\documentclass[pop,fleqn]{w-art}
\usepackage{times}
\usepackage{w-thm}
\usepackage{amsmath}
\usepackage{amsfonts}
\usepackage{amssymb}

\theoremstyle{plain}

\theoremstyle{definition}

\usepackage{graphicx}

\def\one{{\rm 1\kern -.9mm l}}

\begin{document}
\DOIsuffix{theDOIsuffix}
\Volume{51}
\Issue{1}
\Month{01}
\Year{2003}
\pagespan{1}{}
\rightline{FFUOV-09/01; DFTT-32/2009} \vspace{0.25cm}



\title [Update on the quantum properties of the M2]{Update on the quantum properties of the Supermembrane.}


\author[M.P. Garcia del Moral]{M. P. Garc\'\i a del Moral\inst{1},\inst{2}%
  \footnote{ E-mail:~\textsf{garcia@to.infn.it}}}
\address[\inst{1}]{Dip. di Fisica Teorica, Universit\`a di Torino
and I.N.FN., sez. di Torino\\
Via Pietro Giuria 1, 10125 Torino (Italy)}
\address[\inst{2}]{Departamento de F\'{\i}sica, Universidad de Oviedo\\
 Avda Calvo Sotelo 18,3307, Oviedo, Spain} 

\begin{abstract}
 In this note we summarize some of the quantum  properties
 found since the early 80's until nowdays that characterize at quantum level
 the spectrum of the supermembrane. In particular we will focus on a topological
 sector of the 11D supermembrane that, contrary to the general case, has a purely discrete spectrum at supersymmetric level.
  This construction has been consistently implemented in different types of backgrounds: toroidal and orbifold-type
   with G2 structure able to lead to a true G2 compactification manifold. This theory  has $N=1$ supersymmetries in 4D.
    comment on the relevant points of this construction as well as on its spectral characteristics. We will
    also make some comments on the quantum properties of some effective formulation of  multiple M2's theories recently found.
\end{abstract}
\maketitle                   





\section{The 11D Supermembrane.}
 The 11D supermembrane, also called M2 brane, \cite{bst} was
discovered in 1987. A year later  its hamiltonian formulation in the
Light Cone Gauge (L.C.G.) and its matrix regularization by
\cite{dwhn} was obtained. The supermembrane is the natural extension
of the string in 11D and it was thought to be a fundamental object
in 11D.
  However, the spectral properties of the regularized M2 differed substantially
    from those of the string as it was shown in a rigorous proof by \cite{dwln}.  Firstly, the theory of the
    supermembrane is a constrained system highly nonlinear and difficult to solve, in distinction with the
    harmonic oscillator-type hamiltonian of the string. Secondly, the supermembrane classically contains
     instabilities that make  the scalar potential along directions in the configuration space vanish. They are called
     'valleys' and render the classical system unstable. The third and most important difference
     is that its supersymmetric spectrum is continuous. Let us characterize it in more
     detail. The hamiltonian of the $D=11$ Supermembrane \cite{bst} may be
defined in terms of maps $X^{\mu}, \mu=0,\dots, 10$, from a base
manifold $\Sigma\times R$ onto a target manifold which we will
assume to be $11D$ Minkowski. $\Sigma$ is a Riemann surface of genus
$g$.  $\sigma^{a}$, $a=1,2$ are local spatial coordinates over
$\Sigma$ and $\tau\in R$ represents the worldvolume time.
Decomposing $X^{\mu}$ and $\Gamma_{\mu}$
 accordingly to the standard
L.C.G ansatz and solving the constraints, the canonical reduced
hamiltonian to the light-cone gauge has the expression
\begin{equation}\label{e1}
 H=\int_\Sigma  \sqrt{W} \left(\frac{1}{2}
\left(\frac{P_m}{\sqrt{W}}\right)^2 +\frac{1}{4}\sqrt{W}
\{X^m,X^n\}^2+
\sqrt{W}\overline{\theta}\Gamma_{-}\Gamma_{m}\{X^m,\theta\}\right)
\end{equation}
where the range of $m$ is now $m=1,\dots,9$ corresponding to the
transverse coordinates in the light-cone gauge, and
 \begin{equation} \label{e4}\{X^{m}, X^{n}\}= \frac{\epsilon
^{ab}}{\sqrt{W(\sigma)}}\partial_{a}X^{m}\partial_{b}X^{n}.
\end{equation} $W(\sigma)$ is a scalar density introduced in the
light-cone gauge fixing procedure. $\theta$ represents the 11D
Majorana spinors. \newline The theory has a residual symmetry which
is the invariance under the area preserving diffeomorphism so it is
 subject to  the following constraints,
\begin{equation}  \label{e2}
\phi_{1}:=d(\frac{P_{m}}{\sqrt{W}}dX^{m}+\overline{\theta}\Gamma_{-}d\theta)=0\qquad\qquad
\phi_{2}:=
   \oint_{C_{s}}(\frac{P_m}{\sqrt{W}}dX^m+\overline{\theta}\Gamma_{-}d\theta) = 0,
\end{equation}
 where $C_{s}$, $s=1,\dots,2g$ is a basis of  1-dimensional homology on
$\Sigma$.
\begin{itemize}
\item{\bf Classical Analysis: String-like spikes}
\end{itemize}
The scalar potential of the supermembrane vanishes along some configurations  known as String-like spikes.
 They are singular configurations associated to configurations of the bosonic maps which depend
 on a single worldvolume variable (or a combination of the two spatial variables)  $X^{m}(a\sigma^1+b\sigma^2,\tau)$
 that render the classical potential unstable. The system fluctuates along all the allowed states, consequently not preserving neither the
 topology nor the number of particles.  Since the membrane may split and merge without any cost of energy,
 the concept of particle looses its meaning,
 and this fact can be taken as a first indication that the 11D Supermembrane can be seen as
 an effective theory of interacting lower-dimensional objects.
 \begin{itemize}
\item{\bf Bosonic Analysis}
\end{itemize}
At quantum level, however, the situation changes completely and the bosonic hamiltonian has
a purely discrete spectrum in spite of the classical instabilities. Its analysis was performed
in the context of matrix model formulation. The regularized hamiltonian of the M2 brane in the LCG is the following \cite{dwhn}:
\begin{equation}
H=Tr[\frac{1}{2}(P^{mA})^2
+\frac{1}{4}(X^{mB}X^{nC}f_{ABC})^2-\frac{i}{2}f_{ABC}X^{mA}\theta^{B}\gamma_{m}\theta^{C}]\end{equation}
subject to the Gauss constraint
\[
\phi_A=f_{ABC}(X^{mB}P_{m}^{C}-\frac{i}{2}\theta_{\alpha}^B\theta_{\alpha}^C)\approx
0.\] The sufficient condition for the bosonic regularized
hamiltonian was formerly  found by L\"uscher \cite{luscher} and
Simons \cite{simons}. In \cite{inertia}
 we have been able to improve these bounds by finding the necessary and sufficient condition
 for the discreteness of the regularized 11D membrane.
 That condition is explained in detail in \cite{inertia}.
 We showed that then, the most precise
condition ensuring the discreteness of the spectrum of the membrane
theories, is given in terms of an intrinsic moment of inertia of the
membrane. It may be interpreted as if the membrane, or equivalently
the $D0$ branes describing it, have a rotational energy. It is a
quantum mechanical effect. The condition is obtained from the
Molchanov \cite{molch}, Maz'ya and Shubin \cite{mazshu} necessary
and sufficient condition on the potential of a Schr\"odinger
operator to have a discrete spectrum. The criteria is expressed not
in terms of the behaviour of the potentials at each point, but by a
mean value, on the configuration space.  The mean value in the sense
of Molchanov considers the integral of the potential on a finite
region of configuration space. It can be naturally associated to a
discretization of configuration space in the quantum theory. We
found that the mean value in the direction of the valleys where the
potential is zero, at large distances in the configuration space, is
the same as a harmonic oscillator with frequencies given by the
tensor of inertia of the membrane.  The interesting feature is that
all previously known bounds for the membrane and Yang-Mills
potential were linear on the configuration variables, while the
bound we found is quadratic on the configuration variables.
\newline
Since the dimensionally reduced action of SYM to $0+1$D corresponds exactly
to the regularized action of the 11D Supermembrane, as indicated in \cite{dwhn},
then the previous result also holds for the bosonic YM in this regime (Slow-mode),
both interpreted as the interaction of multiple D0 branes. The Slow-mode regime assumes that
there is no momentum for the $D0$ branes but they still have energy. In \cite{inertia}
we also obtained analytically bounds, based on the moment of inertia for the
hamiltonian of  $0+1$D Yang-Mills theories which
allows to obtain interpretation about the mass gap of the theory.
In particular we found that a lower bound for $SU(3)$ in $3+1$ dimensions is given by
a hamiltonian whose spectrum and eigenvalues are known, and its
eigenfunctions are expressed in terms of Bessel functions.
\begin{itemize}
\item{\bf Supersymmetric Analysis}
\end{itemize}
A quantum mechanic supersymmetric hamiltonian is realized
in a matrix whose diagonal contains the bosonic part of the hamiltonian while the
 fermionic potential lies in the non-diagonal entries.
\begin{equation*}
H=-\Delta+V_{B}\mathbb{I}+V_{F}
\end{equation*}
We denote by $H$ the hamiltonian operator defined in the whole phase
space although the hamiltonian operator if the 11D Supermembrane is
defined only in an hypercone which is a subset of the previous space
of those solutions that satisfy the first class constraints of the
supermembrane. The core of the proof, which was done in \cite{dwln},
consisted in the construction of a wavefunction such that for any
given energy $E\ge 0$ there exists a suitable wave function $ \Psi$
with $||\Psi||=1$ such that for any $\epsilon>0$ is always satisfied
\begin{equation}
||(H-E)\Psi||<\epsilon.
\end{equation}
The $SU(N)$ regularized model obtained from (\ref{e1}) \cite{dwhn}
was shown to have conti\-nuous spectrum from $[0,\infty)$,
\cite{dwln},\cite{dwmn},\cite{dwhn}. It happens that the fermionic
contribution cancel this effect of raising the valleys and lead to a
continuous spectrum at quantum level. The supermembrane in 11D could
not be considered any longer as a fundamental object, contrarily it
was interpreted as a second quantized theory, a theory of
interacting $D0$'s. In 1996, \cite{bfss} conjectured that the action
that should be taken as a fundamental in the context of M-theory
should be the matrix model action of $D0$ branes (related with the
IKKT formulation in terms of $D(-1)$ branes). This led to the
fruitful field of matrix model formulation so much developed since
then. Almost simultaneously, there was an study on the supermembrane
with winding started by \cite{dwpp} in a series of papers, where
they extended the study to compactified spaces and were able to see
that compactification by itself is not able to avoid the string-like
spikes that together with supersymmetry render the spectrum
continuous. Fortunately a new avenue was explored: the discovering
of topological sector inside the supermembrane that has a purely
discrete spectrum: The MIM2.

\section{A type of Supermembranes with discrete spectrum: The MIM2}
In what follows we will consider a topological restriction on the
configuration space of the 11D supermembrane. This topological
sector is characterized by imposing an irreducibility winding
condition on the compactified target space.  It generates a central
charge in the supersymmetric algebra of the 11D supermembrane.
Geometrically it corresponds to a Supermembrane minimally immersed
in the target space. For that reason from now on, we will refer to
it as MIM2, \cite{gmr}-\cite{bgmr2}. Following \cite{bellorin} we
may extend the original construction on a $M_{9}\times T^{2}$ to
$M_{7}\times T^{4}$, $M_{5}\times T^{6}$ target manifolds by
considering genus $1,2,3$ Riemann surfaces on the base respectively.
We are interested in reducing the theory to a 4 dimensional model,
we will then assume a target manifold $M_{4}\times T^{6}\times
S^{1}$. The configuration maps satisfy:
\begin{equation}\label{e5}
 \oint_{c_{s}}dX^{r}=2\pi S_{s}^{r}R^{r}\quad r,s=1,\dots,6, \quad
 \oint_{c_{s}}dX^{m}=0 \quad m=8,9; \quad \oint_{c_{s}}dX^{7}=2\pi L_{s}R,
 \end{equation}
where $S^{r}_{s}, L_{s}$ are integers and $R^{r}, r=1,\dots,6$ are
the radius of $T^{6}=S^{1}\times\dots\times S^{1}$ while $R$ is the
radius of the remaining $S^{1}$ on the target.
We now impose the central charge condition \begin{equation}\label{e8}
I^{rs}\equiv \int_{\Sigma}dX^{r}\wedge dX^{s}=(2\pi
R^{r}R^{s})\omega^{rs}n \end{equation} where $\omega^{rs}$ is a symplectic
matrix on the $T^{6}$ sector of the target and $n$
denotes an integer representing the irreducible winding. The topological
condition (\ref{e8}) does not change the field equations of the
hamiltonian (\ref{e1}). In addition to the field equations obtained
from (\ref{e1}), the classical configurations must satisfy the
condition (\ref{e8}). In the quantum theory, the space of physical
configurations is also restricted by the condition
(\ref{e8}) \cite{torrealba},\cite{ovalle}.\\
We consider now a Hodge decomposition of the map satisfying
condition (\ref{e8}): \begin{equation}
dX^{r}=M_{s}^{r}d\widehat{X}^{s}+dA^{r}
\end{equation} where $d\widehat{X}^{s}$, $s=1,\dots,2g$ is a basis
of harmonic one-forms over $\Sigma$, $M_{s}^{r}$ is  a matrix of
integers carrying the d.o.f of harmonic forms and and $dA^{r}$
represents the single-valued one-forms. Now, we impose the
constraints (\ref{e2}). It turns out that $M_{s}^{r}$ gets fixed and
can be expressed in terms of a matrix $S\in
Sp(2g,Z)$,\cite{joselen}. The natural choice for $\sqrt{W(\sigma)}$
in this geometrical setting is define
$\sqrt{W(\sigma)}=\frac{1}{2}\partial_{a}\widehat{X}^{r}\partial_{b}\widehat{X}^{s}\omega_{rs}.$
$\sqrt{W(\sigma)}$ is then invariant under the change $
d\widehat{X}^{r}\to S_{s}^{r}d\widehat{X}^{s}$. We thus conclude
that the theory  is invariant not only under the diffeomorphisms
generated by $\phi_{1}$ and $\phi_{2}$ but also under the
diffeomorphisms, biholomorphic maps, changing the canonical basis of
homology by a modular transformation. The theory of supermembranes
with central charges in the light cone gauge (LCG) we have
constructed depends then on the moduli space of compact Riemanian
surfaces $M_{g}$ only. In addition, when compactified to 9D there
has been proved in \cite{gmmr} that the hamiltonian is also
invariant under a second SL(2,Z) symmetry associated to the $T^{2}$
target space that transform
 the Teichm\"uller parameter of the 2-torus $T^{2}$.

This construction can be seen in detailed in \cite{joselen}. There
the theory is formulated in 4D and its hamiltonian is the following:
\begin{equation}\label{e}
\begin{aligned}
H_{d}=&\int \sqrt{w}d\sigma^{1}\wedge d\sigma^{2}[\frac{1}{2}(\frac{P_{m}}{\sqrt{W}})^{2}
+\frac{1}{2}(\frac{\Pi^{r}}{\sqrt{W}})^{2}+\frac{1}{4}\{X^{m},X^{n}\}^{2}+\frac{1}{2}(\mathcal{D}_{r}X^{m})^{2}\\
& \nonumber +\frac{1}{4}(\mathcal{F}_{rs})^{2}+\frac{1}{2}(F_{ab}\frac{\epsilon^{ab}}{\sqrt{W}})^{2}
+\frac{1}{8}(\frac{\Pi^{c}}{\sqrt{W}}\partial_{c}X^{m})^{2}+\frac{1}{8}[\Pi^{c}\partial_{c}(\widehat{X}_{r}+A_{r})]^{2}]+\\
& \nonumber \Lambda(\{\frac{P_{m}}{\sqrt{W}},X^{m}\}-\mathcal{D}_{r}\Pi^{r}
-\frac{1}{2}\Pi^{c}\partial_{c}(F_{ab}\frac{\epsilon^{ab}}{\sqrt{W}}))+\lambda\partial_{c}\Pi^{c}]+ \\
\nonumber &+\int_{\Sigma} \sqrt{W}[-\overline{\Psi}\Gamma_{-}\Gamma_{r}\mathcal{D}_{r}\Psi
+ \overline \Gamma_{-}\Gamma_{m}\{X^{m},\Psi\}+1/2\overline{\Psi}\Gamma_{7}\Pi^{b}\partial_{b}\Psi]+\Lambda \{\overline{\Psi}\Gamma_{-}, \Psi\}
\end{aligned}
\end{equation}
where  $\mathcal{D}_r X^{m}=D_{r}X^{m} +\{A_{r},X^{m}\}$,
$\mathcal{F}_{rs}=D_{r}A_s-D_{s }A_r+ \{A_r,A_s\}$, \\
 $D_{r}=2\pi
R^{r}\frac{\epsilon^{ab}}{\sqrt{W}}\partial_{a}\widehat{X}^{r}\partial_{b}$
and $P_{m}$ and $\Pi_{r}$ are the conjugate momenta to $X^{m}$ and
$A_{r}$ respectively. $\Psi$ are  $SO(7)$ Majorana spinors.
$\mathcal{D}_{r}$ and $\mathcal{F}_{rs}$ are the covariant
derivative and curvature of a symplectic noncommutative theory
\cite{ovalle},\cite{bgmmr}, constructed from the symplectic
structure $\frac{\epsilon^{ab}}{\sqrt{W}}$ introduced by the central
charge. The integral of the curvature we take it to be constant and
the volume term corresponds to the value of the hamiltonian at its
ground state. The physical degrees of the theory are the $X^{m},
A_{r},\Pi^{c}, \Psi$. They are single valued fields on $\Sigma$. It
was shown in \cite{joselen} that the supermembrane minimally immerse
on this $T^7$ has no moduli free for the isotropic tori, is $N=1$ (4
target space supersymmetries) and has been proved in \cite{joselen}
that it has a purely discrete spectrum  with finite multiplicity and
compact resolvent following the lines of the theorems shown in
Section 3. Moreover it is possible to define the
 supermembrane with central charges induced through an irreducible winding on
 an orbifold with G2 structure and define precisely all of the maps of the twisted states.
 Since the symmetries of the orbifold are symmetries of the parent theory in the untwisted sector,
 then no state is projected out and it coincides with those of the 4D formulation of the MIM2 on the $T^7$. Due to
the regularized parent hamiltonian symmetries one can guaranteed
that the spectral properties are inherited on the orbifolded theory
since the twisted sector only adds a finite number of states that do
not change the qualitative properties of the spectrum. Once that the
singularities are resolved by the action of the symplectomorphisms
present in the theory one ends with the well-known example of
compact G2 manifold with the supermembrane minimally immersed on it.

\section {Discreteness of the spectrum}
In this section I would like to illustrate the master lines followed
to show the discreteness of this sector of the supermembrane theory.
The analysis of the spectrum of the truncated Schr\"odinger
operator associated to $\widehat{H}$ without further requirements on the constants $f^{N}$:\\
 i) The potential of the Schr\"odinger  operator only vanishes at
the origin of the configuration space:
\begin{equation} V=0 \to \vert\vert (X,A,\phi)\vert\vert=0 \end{equation}
where $\vert\vert .\vert\vert$ denotes the euclidean norm in
$R^{L}$. We notice that the original hamiltonian as well as
$\widehat{H}$ are defined on fields up to constants. This condition
guarantee the non-existence of singular configuration in the
hamiltonian.\newline ii) There exists a constant $M> 0$ such that
\begin{equation} V(X,A,\phi)\ge M\vert\vert (X,A,\phi))\vert\vert^{2}.\end{equation}
Again, this bound arises from very general considerations. In fact,
writing $(X,A,\phi)$ in polar coordinates $X=Rx\quad A=Ra\quad
\phi=R\varphi$
 where $\theta\equiv (x,a,\varphi)$ is defined on the unit sphere, $\vert\vert(x,a,\varphi)\vert\vert=1$,
\begin{equation}
V(X,A,\phi)\ge M R^{2}.
\end{equation}
The Schr\"odinger operator is then bounded by below by an harmonic
oscillator and goes to infinity on every direction of the
configuration space \cite{simons}. Consequently it has a compact
resolvent. The result coincides qualitatively with the bosonic
statement of discreteness for the regularized membrane without the
central charge condition, however, the bound is not the same. It
depends indirectly on the central charge condition which is
responsible for the generation of the mass terms. The proof of
discreteness for the bosonic part of the supermembrane with central
charges without regularization was done in \cite{bgmr2}. It
corresponds to have infinite d.o.f., then spectral theorems valid
for finite Hilbert spaces generically do not hold.
\begin{itemize}
\item{\bf The Supersymmetric Analysis.}
\end{itemize}
The supersymmetric extension can be proved in different ways \cite{bgmr,br} to be also discrete.
See for example the guide lines of the proof in \cite{bgmr}. By using the Lemma.1 of \cite{bgmr} that states:
\begin{lemma} \label{t1} Let $v_k(x)$ be the
eigenvalues of $V(x)$.
If all $v_{k}(x)\to +\infty$ as $|x|\to \infty$, then the spectrum
of $H$ is discrete.
\end{lemma}
To this end the authors of \cite{bgmr}, decompose the resolvent of
the following operator $\mu$ as
\begin{equation*}
\mu=-\Delta+V_{B}\mathbb{I}+V_{F}
\end{equation*}
where $V_B$ and $V_{F}$ denote the bosonic and fermionic potentials
respectively on the whole space of configurations. The space of
solutions of the supermembrane is smaller since it is constrained.
Then $V_{F}$ is the sum of a linear homogeneous part
$M(X,\mathcal{A})$ and a constant matrix $C$ that may be reabsorbed,
(see \cite{bgmr} for further details). The eigenvalues are
determined by the solutions of the characteristic equation. By
virtue of the homogeneity of $M$, $\lambda$ must satisfy
\begin{equation*}
\det \left[
\frac{\lambda-V_{B}}{R}\mathbb{I}-M(\phi,\psi)\right]=0, \qquad R>0.
\end{equation*}
Therefore if $\widehat{\lambda}$ are the eigenvalues of
$M(\phi,\psi)$, then
\begin{equation*}
\lambda=V_{B}(R\phi,R\psi)+R\widehat{\lambda}.
\end{equation*}
Consequently, $\lambda\to +\infty$ whenever $R\to\infty$. Notice
that $V$ is discrete, hence it is automatically bounded from below.
 Since this is true for the operator $\mu $  whose domain is the whole configuration
 space without considering the constraints, then it also holds for the constrained theory.
  This statement obviously would not hold in the other way-round. Finally it was shown that the
  fermionic contribution to the susy hamiltonian
do not change the qualitative properties of the spectrum of the
hamiltonian. In fact, both contributions are linear on the
configuration variables. In addition the supersymmetric contribution
cancels the zero point energy of the bosonic oscillators even in the
exact theory \cite{stelle}, \cite{bgmr2}.
The Schr\"oedinger operator is then bounded by an harmonic
oscillator. Consequently it has a compact resolvent. We now use
theorem 2 \cite{br} to show that: i) The ghost and antighost
contributions to the effective action assuming a gauge fixing condition linear on the configuration variables, ii) the fermionic
contribution to the susy hamiltonian, do not change the qualitative
properties of the spectrum of the hamiltonian.
\newline
\begin{itemize}
\item{\bf Generic Supersymmetric matrix potential Analysis}
\end{itemize}
Whenever no string-like configurations are present the following
sufficient condition for discreteness hold. The assumptions on the bosonic potential
are very mild, we only require the potential to be measurable,
bounded from below and unbounded above in every direction (if the
potential is continuous the unbounded assumption ensures the
bounded-ness from below). For instance, for a quantum mechanical
potential of the form
\begin{equation}
  V=V_B(x)\mathbb{I}+V_F(x) \in \mathbb{C}^{2^n\times 2^n},
\end{equation}
 where $x\in \mathbb{R}^L$, $V_B(x)$ is continuous with the asymptotic
behaviour
\[
   V_B \geq c \|x\|^{2p}, \qquad c>0,
\]
and the fermionic matrix potential satisfies
\[
 V_F\leq V_F|_{\|x\|=1} \|x\|^q,
\]
for all $\|x\|>R_0$ with $2p>q$, the Hamiltonian of the quantum
system has spectrum consisting exclusively of isolated eigenvalues
of finite multiplicity.
\begin{itemize}
\item{\bf Comments on the multiple M2's spectrum}
\end{itemize}
The discreteness of the spectrum of the bosonic M5 brane
\cite{alvaro} has been characterized following the necessary and
sufficiency condition showed in Section 1 \cite{inertia}. More
generally, the authors have proved that provided a suitable matrix
regularization of a generic p-brane  whose regularized hamiltonian
is of the form,\begin{equation}
    H_{L}= -\Delta+V_{L}(X)=-\Delta+(X^{a_{1}}_{M_{1}}\ldots
    X^{a_{L}}_{M_{L}}f_{a_{1}\ldots a_{L}}^{b})^{2}
\end{equation}
where $L$ is the degree of the brane considered, $M_{i}=
1,\ldots,K$, $a_{i}=1,\ldots,N$, $K\ge L$, $N\ge L$, $X=X^{a}_{M}
\in \mathcal{R}^{KN}$ and $f_{a_{1}\ldots a_{L}}^{b}$ is a
non-singular constant tensor totally antisymmetric and it is not
singular, then the spectrum of the bosonic hamiltonian is discrete
\cite{alvaro}. This condition also holds for the type of bosonic
scalar potentials of the multiple M2's actions based on 3-algebras
(whenever a proper discretization is provided). For example:
\cite{BL2},\cite{gomis}, \cite{schwarz2}. In the case of \cite{abjm}
their action is expressed in terms of an ordinary lie algebra whose
structure constants can be seen as some particular choice of the
3-algebra structure constants that render it not fully
antisymmetric, then the above criteria does not hold. One may
consider a toy model whose bosonic potential is a polynomial product
of an arbitrary number of operators: \[ V(x)=\Pi_{k=1}^n
|x_k|^{\alpha_k},\] where $\alpha_k>0$ for all $k=1,2,\ldots,n$.
Then the spectrum of the Schr\"odinger operator $-\Delta+V$ in
$L^2(\mathbb{R}^n)$ is discrete. This toy model gives a hint that
also the bosonic potential of the $N=6$ Chern-Simons term coupled to
matter will probably also be discrete, although a rigorous study of
its quantum properties is needed to establish it properly
\cite{m2s}. An interesting issue is the spectral characterization of
the complete hamiltonians including their supersymmetric extension.
The analysis is much more involved. All of these actions of multiple
M2's have in common the construction of a conformal supersymmetric
gauge theory, with a sextic scalar potential, quadratic couplings in
the fermionic variables and two-coupled Chern-Simons terms with a
number of supersymmetries, $N=8$ \cite{BL2},\cite{gomis},
\cite{schwarz2} or $N=6$ susy in the \cite{abjm} case. A rigorous
study if its spectral properties is currently under study
\cite{m2s}, here I will just make some heuristical comments. Their
fermionic contribution in distinction with the case of a single M2
brane depends quadratically on the bosonic variables. The sufficient
condition for discreteness of supersymmetric potentials shown
previously is no longer applicable and although it does not exclude
completely the possibility of the spectrum be discrete, makes it
much more fine tuned. On the other hand, the continuity of the
spectrum for a single M2 in \cite{dwln} lies on the result that
along the singularities the potential behaves as a susy harmonic
oscillator so there is no confining potential. The present
dependence of the fermionic terms on the bosonic variables makes it
much more involved and to determine this a much more exhaustive
study is required \cite{m2s} and lies out the scope of this work.
\section{Acknowledgements}
MPGM thanks the organizers of the Workshop ``ForcesUniverse 2008''
celebrated in Varna, Bulgaria for the possibility to present this
work. MPGM also would like to thank to M. Dimitrijevic, J.M. Pe\~na
and A. Restuccia for their comments on the manuscript. MPGM is
partially supported by the European Comunity's Human Potential
Programme under contract MRTN-CT-2004-005104 and by the Italian MUR
under contracts PRIN-2005023102 and PRIN-2005024045 and by the
Spanish Ministerio de Ciencia e Innovaci\'on (FPA2006-09199) and the
Consolider-Ingenio 2010 Programme CPAN (CSD2007-00042).

\end{document}